\documentclass[aps,pra,onecolumn,superscriptaddress,notitlepage]{revtex4-2}
\usepackage{tikz}
\usetikzlibrary{calc}
\usetikzlibrary{shapes.geometric, arrows}
\usepackage{eufrak}
\usepackage{ulem}
\usepackage{enumitem}
\usepackage{mathtools}
\usepackage{graphicx}
\usepackage{subcaption}
\usepackage{adjustbox} 
\usepackage{caption}
\usepackage{siunitx}
\usepackage{booktabs}

\usetikzlibrary{positioning}
\usepackage{bm}

\usepackage{xurl}                 
\usepackage[most]{tcolorbox}
\usepackage{soul} 
\usepackage[]{xcolor}
\sethlcolor{blue!20}

\newcommand{\bematrix}{\left(\begin{matrix}}
\newcommand{\ematrix}{\end{matrix}\right)}

\usepackage{ulem}
\usepackage{dcolumn}
\usepackage{amsmath,amssymb}
\usepackage{bm}
\usepackage{bbm}
\usepackage{overpic}
\usepackage{latexsym}
\usepackage{color}
\usepackage[english]{babel}
\usepackage{latexsym}
\usepackage{psfrag,graphicx}
\usepackage{epsf}
\usepackage{amsmath}
\usepackage{amssymb}
\usepackage{amsfonts}
\usepackage{natbib}
\usepackage{multirow} 
\usepackage{appendix}
\usepackage{verbatim}
\usepackage{enumitem}
\usepackage{dsfont}
\usepackage{float}
\usepackage{tikz}

\definecolor{mygrey}{gray}{0.35}
\definecolor{myblue}{rgb}{0.2,0.2,0.8}
\definecolor{myzard}{cmyk}{0,0,0.05,0}
\definecolor{mywhite}{rgb}{1,1,1}
\definecolor{myred}{rgb}{0.9,0.1,0.}
\usepackage[colorlinks=true,citecolor=myblue,linkcolor=myblue,urlcolor=myblue]{hyperref}

\usepackage[makeroom]{cancel}

\newenvironment{proof-of}[1]{\medskip\noindent\textbf{Proof of {#1}.}}{\hfill$\blacksquare$\medskip}

\usepackage{colortbl}
\usepackage{xcolor}

\usepackage{tcolorbox}

\date{}
\begin{document}

\title{Deploying and validating a metropolitan QKD secure network: architecture and field performance}

\author{Claudia De Lazzari}
\author{Nicola Biagi}
\author{Damiano Giani}
\affiliation{QTI S.r.l. --- Largo Enrico Fermi 6, 50125, Florence, Italy}

\author{Marco Russo}
\affiliation{Politecnico di Torino --- Corso Duca degli Abruzzi, 24, 10129 Turin, Italy}

\author{Fernando Chirici}
\affiliation{QTI S.r.l. --- Largo Enrico Fermi 6, 50125, Florence, Italy}

\author{Francesco Stocco}
\affiliation{Telsy S.p.A. --- Corso Svizzera 185, 10149, Turin, Italy}

\author{Saverio Francesconi}
\author{Giacomo Ferranti}
\affiliation{QTI S.r.l. --- Largo Enrico Fermi 6, 50125, Florence, Italy}

\author{Alessandro Soureal}
\author{Antonella Sanguineti}
\affiliation{TI Sparkle S.p.A. --- Via di Macchia Palocco 223, 00125 Rome, Italy}

\author{Bartolomeo Montrucchio}
\affiliation{Politecnico di Torino --- Corso Duca degli Abruzzi, 24, 10129 Turin, Italy}

\author{Christian Laurenzi}
\author{Oliviero Testa}
\author{Guglielmo Morgari}
\affiliation{Telsy S.p.A. --- Corso Svizzera 185, 10149, Turin, Italy}

\author{Antonio Manzalini}
\affiliation{TIM S.p.A. --- Via Gaetano Negri, 1, 20123 Milan, Italy}

\author{Tommaso Occhipinti}
\affiliation{QTI S.r.l. --- Largo Enrico Fermi 6, 50125, Florence, Italy}

\author{Alessandro Zavatta}
\affiliation{QTI S.r.l. --- Largo Enrico Fermi 6, 50125, Florence, Italy}
\affiliation{Istituto Nazionale di Ottica - Consiglio Nazionale delle Ricerche (INO-CNR) --- Largo Enrico Fermi 6, 50125 Florence, Italy}

\author{Davide Bacco}
\email{davide.bacco@unifi.it}
\affiliation{QTI S.r.l. --- Largo Enrico Fermi 6, 50125, Florence, Italy}
\affiliation{Department of Physics and Astronomy, University of Florence --- 50019, Florence, Italy}

\begin{abstract}

The advent of cryptographically relevant quantum computers poses an existential threat to classical public-key infrastructure. Quantum Key Distribution (QKD) addresses this challenge by providing information-theoretic security for key establishment, independently of any computational hardness assumption. In this work, the deployment and experimental validation of a metropolitan-scale quantum-secure network between data centers in Milan is reported. The network operates over installed fiber infrastructure and implements a layered architecture integrating QKD hardware, standards-compliant Key Management (KM), and centralized Software-Defined Networking (SDN) orchestration. Dynamic path reconfiguration via active optical switching and trusted-node routing allow automated fail-over solutions. Application-layer validation across diverse protocols and workloads confirms the seamless interoperability of all system components. These results establish the technical and operational readiness of metropolitan QKD networks for production deployment, and offer a replicable blueprint for building quantum-secure communication infrastructure at metropolitan scale.\\

Keywords: Quantum Key Distribution (QKD), Metropolitan Area Network (MAN), Star topology network, Software defined networking, Data center, Quantum-secure infrastructure, Trusted node.

\end{abstract}

\makeatletter
\let\@date\@empty 
\makeatother

\maketitle

\section{Introduction}
With the emergence of the quantum computing threat, which poses a significant risk to current encryption methods used to secure information transmission, the global community is preparing for a transition toward quantum-resistant solutions. While new classical approaches are under development, forming the foundation of post-quantum cryptography, quantum physics itself has enabled the design of an unconditionally secure method for symmetric key distribution, known as quantum key distribution (QKD). 
Unconditional security is fundamentally grounded in quantum mechanics and guarantees security against any physically admissible attack, classical or quantum.
Since the first experimental realization of QKD \cite{Brassard1989Dawn}, technological development has focused on extending achievable distances \cite{Lucamarini2018Overcoming} and enabling new architectures, including satellite-based key distribution \cite{Liao2017Satellite}.

National and international programs have begun deploying QKD infrastructures at scale. For instance, the European Union is rolling out the European Quantum Communication Infrastructure (EuroQCI), a continental initiative involving all 27 Member States to establish a secure quantum communication network composed of both terrestrial fiber-based links and a satellite segment. 

As quantum communication evolves toward larger and increasingly complex network infrastructures, robust coordination mechanisms become essential to ensure secure and reliable operation across heterogeneous environments. To meet these requirements, the architecture integrates standards-compliant Key Management (KM) and a centralized Software-Defined Networking (SDN) control plane, enabling unified orchestration, automated configuration, and efficient resource utilization \cite{Martin2024MadQCI}.
In particular, the KM layer exposes standardized interfaces for application key delivery, aligned with ETSI GS QKD 014 \cite{ETSIGSQKD014}, to ensure interoperability across the security stack.

It is within this strategic framework that the Metropolitan Area Network (MAN) presented in this work is situated. This deployment contributes directly to the broader European effort to establish a secure, large-scale quantum communication infrastructure. 

The manuscript presents the final validation of a QKD network field trial in the metropolitan area of Milan, conducted over a commercial fiber infrastructure connecting three data centers. The paper describes the architecture and topology of the QKD network, the orchestration components, and the experimental tests and results. 
The focus spans the complete stack -- from quantum key generation through key management services and up to application-layer consumption -- to assess interoperability, operational robustness, and readiness for real-world integration. 
 
We summarize overall performance in terms of Secure Key Rate (SKR), within the constraints of deployed routes and equipment, and we report interoperability results across layers.
Operational behavior under threshold-based activation is reached by configuring the system to maintain a target inventory of cryptographic keys at each QKD node. 
This result aligns with the intended orchestration policies and provides a clear basis for tuning activation and replenishment strategies.

The trial confirms the feasibility of stable secure key generation under realistic metropolitan conditions and demonstrates correct operation of key management functions, including distribution, storage, retrieval, and life-cycle control via interoperable interfaces. 
Applications successfully consumed QKD-derived keys, validating cryptographic workflows and evidencing functional readiness for selected use cases, and further demonstrating the maturity of the overall integration of the QKD layer in complex networks and existing infrastructures. 

\section{Network configuration}
\label{sec:network}
The objective of this field deployment is to demonstrate a functional, dynamically reconfigurable infrastructure that delivers end-to-end data traffic encrypted with symmetric keys supplied by QKD systems in real time, while operating over a complex, time-varying topology.
To characterize the MAN, we begin with the physical topology, namely, the locations grounded in existing infrastructure, and the interconnecting fiber plant. We then introduce the network logical layers and discuss how they are overlaid on this physical substrate.

\label{subsec:layers}
\begin{figure}[ht]
    \centering
    \includegraphics[width=1\linewidth]{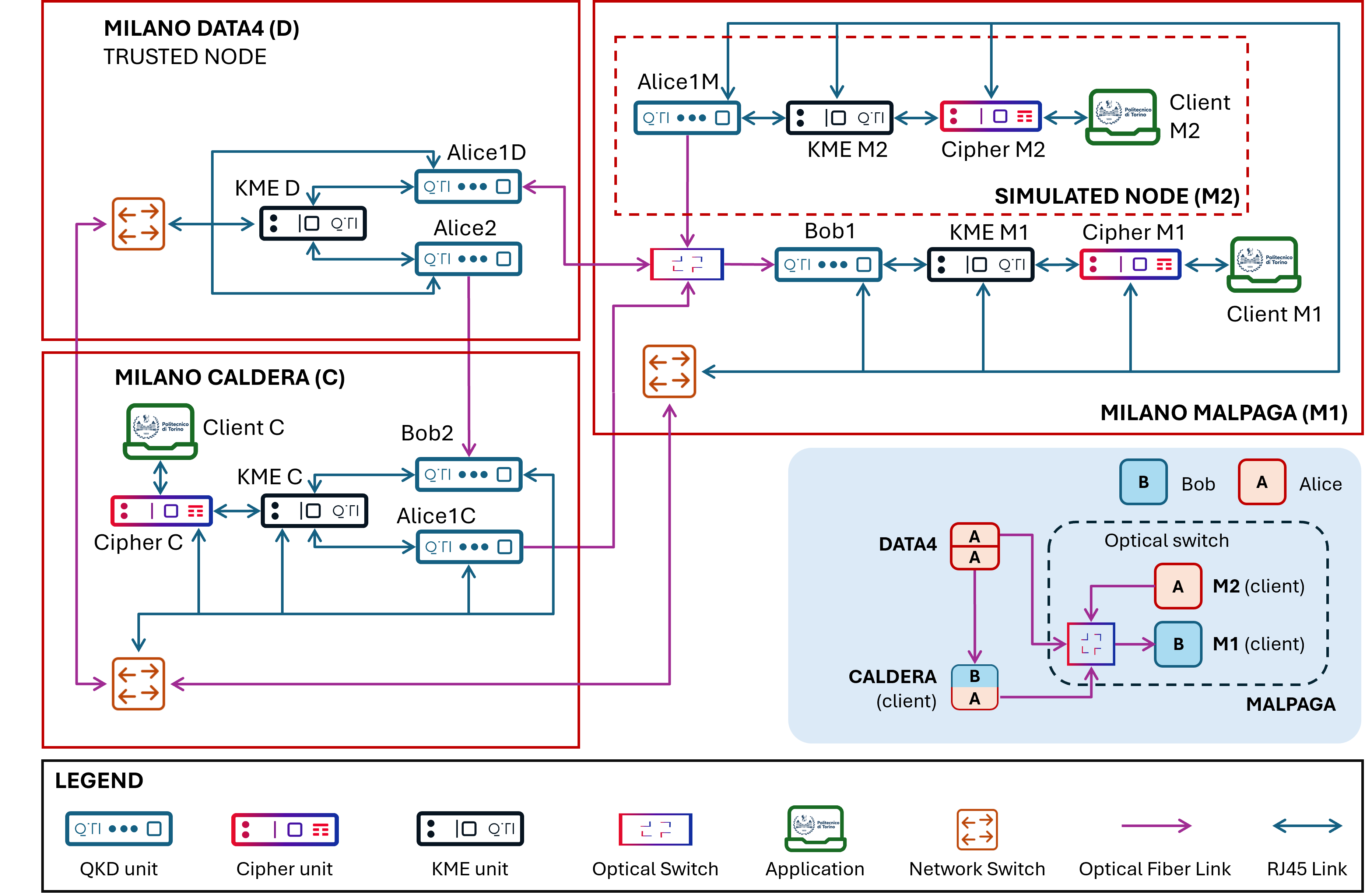}
    \caption{Network architecture overview. The three red boxes represent the Points of Presence (PoPs): Milano Data4 (node D), Milano Caldera (node C), and Milano Malpaga. The latter PoP hosts two nodes — M1 and M2 — with the second implemented using a fiber spool for simulation purposes. The inset at the bottom right shows a high-level representation of the node configuration. Refer to the main text for a detailed description of the network.}
    \label{fig:network}
\end{figure}

The network is a four-node topology for QKD deployment across the metropolitan area of Milan, interconnecting three physical locations, or point-of-presence (PoPs), represented in Fig.~\ref{fig:network}:
\begin{enumerate}
    \item PoP Milano Data4 (Campus) --- {Via Monzoro, Cornaredo 20007 (MI), Italy};
    \item PoP Milano Malpaga (TI Sparkle) --- {Via Strada Antica di Cassano,
20051 Cassina de' Pecchi (MI),
Italy};
\item PoP Milano Caldera --- {Via Caldera, 20153 (MI), Italy}.
\end{enumerate}

Importantly, all PoPs are situated within fully operational commercial data-center environments, where the QKD nodes coexist with production ICT systems and facility constraints, ensuring that the deployment reflects realistic conditions of metropolitan-scale telecom networks rather than a controlled laboratory setup.

The primary inter-PoP fibers are G.652 single-mode (SM) dark fiber pairs, suitable for QKD quantum channels and the associated service channels. The corresponding lengths and losses are reported in Table~\ref{tab:links}, whose values correspond to optical-time-domain reflectometer {(OTDR)} measurement results obtained between the different PoPs. The reported attenuation values were measured at 1550 nm and do not include losses introduced by equipment inside the PoPs.\\

\begin{table}[ht]
    \centering
    \begin{tabular}{lcc}
        \toprule
        \textbf{Link} & \textbf{Fiber length (km)} & \textbf{Loss (dB)} \\
        \midrule
        Data4 -- Malpaga   & 50.965 & 13   \\
        Malpaga -- Caldera & 37.793 & 9.9  \\
        Caldera -- Data4   & 8.678  & 2.6  \\
        \bottomrule
    \end{tabular}
    \caption{Links, lengths and losses of the G.652 dark fibers connecting the three locations.}
    \label{tab:links}
\end{table}

The network topology forms a closed triangular metropolitan link enabling multi-hop or point-to-point QKD operation across all three PoPs.
In order to implement the four-node topology, a simulated external node is configured inside PoP Milano Malpaga (Malpaga M2), equipped with a {40 km} fiber spool serving as an emulated channel between the simulated external QKD transmitter unit and the on-site receiver setup. 

\subsection{Network architecture}

Delivering an application that exchanges data traffic secured by QKD-generated keys requires the introduction of four network logical layers, which we enumerate bottom-up (from lowest to highest) for clarity:
\begin{enumerate}
    \item {[QKD]} The Quantum Key Distribution layer, which generates symmetric cryptographic keys along each network topology path.
    \item {[KM]} The Key Management layer, composed by Key Management Entities (KMEs), and responsible for storing and distributing the keys. KMEs act also as SDN controller and agents, ensuring dynamic network control and reconfiguration.
    \item {[ENC]} The encryption  layer, composed by Ciphers (encryptors and decriptors), which use the quantum-generated keys to encrypt/decrypt the traffic.
    \item {[APP]} The Applicative  layer, composed by Clients (Cs) that generate the data traffic to be secured.
\end{enumerate}
The devices are distributed across the three different locations, as depicted in Fig.~\ref{fig:network}, and detailed below. 
The providers of the layers are QTI for QKD and KM, Telsy for ENC and Politecnico di Torino for APP layer. 

The pairs of single-mode fibers that interconnect the PoPs serve two functions. 
Each node is equipped with a network switch that features small form-factor plug-in (SFP) transceiver ports. By using dense-wavelength-division-multiplexing (DWDM) MUX/DEMUX devices, the transmitted and received signals from each transceiver are multiplexed and forwarded to the next node using the first fiber of the pair. This design ensures full mesh connectivity for all devices while using just a single fiber. All devices within each node are connected to the local switch via RJ45 cables.

The second fiber carries the signals required for quantum states transmission and module synchronization, as described in the following paragraph.

\subsubsection{Quantum layer}
\label{par:qkd}
A QKD system is mainly composed by two units, a transmitter and a receiver. The QKD protocol prescribes a quantum states exchange stage and a classical post-processing stage, in order to generate the final secure and symmetric key. Transmitter and receiver are usually named Alice and Bob, respectively. In the implemented MAN we installed the following QKD transmitters: 
    \begin{enumerate}
        \item Alice1C in PoP Milano Caldera;
        \item Alice1M in PoP Milano Malpaga;
        \item Alice1D in PoP Milano Data4.
    \end{enumerate}
These three transmitters can be routed to the Bob1 receiver, situated in the Milano Malpaga PoP, by appropriately configuring the optical switch located within that Po
A fourth transmitter, Alice2, is deployed at the Milano Data4 PoP and is directly linked to the Bob2 receiver in the PoP Milano Caldera, providing a stable point-to-point QKD connection. \\

The QKD devices used in this setup are QuellX systems from QTI, engineered to enable the simultaneous transmission of quantum and classical signals over a single optical fiber. In this configuration, quantum states are transmitted in the O-band (at 1310 nm), while synchronization signals are carried in the C-band (at 1550 nm). Carefully designed optical filters minimize Raman scattering and enhance the overall signal-to-noise ratio, ensuring stable and efficient quantum key distribution.

\subsubsection{Key management layer}
\label{par:kms}
The KM system is a network appliance designed to unify secure key storage, key distribution, and key life-cycle management across the quantum-secure infrastructure. It operates as a flexible, modular platform where multiple services collaborate to receive, manage and monitor keys, store them safely, and deliver them to applications via standardized interfaces. The KM layer is built around QTI’s Quantum KME (QKME) appliances.  

From an architectural perspective, the number of KMEs -- fundamental units of the KM system -- can be optimized by strategically deploying one unit at each physical node. In practice, in this MAN topology: 
\begin{enumerate}
    \item KME C is installed at PoP Milano Caldera to manage keys genereted by Bob2 and by Alice1C;
    \item KME D at PoP Milano Data4 manages Alice1D and Alice2;
    \item two separate KMEs are installed at PoP Milano Malpaga, KME M2 coupled to Alice1M and KME M1 to Bob1, respectively.
\end{enumerate}   
The apparent redundancy in the Malpaga PoP is explained by the objective of simulating an external node (M2), as shown in Fig.~\ref{fig:network}.

The KM layer also benefits from centralized Software-Defined Networking (SDN) orchestration. In this architecture, KME C, KME D and KME M2 operate as network agents, while KME M1 functions as the SDN controller. This setup enables continuous monitoring of network conditions and allows dynamic reconfiguration of optical paths through an optical switch in the event of a failure, ensuring adaptive and resilient network topology management.

\subsubsection{Encryption and Applicative layers}
\label{par:enc}
Quantum-generated keys are consumed at the ENC layer to provide confidentiality and integrity protection for application traffic. 
The ciphers, MusaX provided by Telsy, are responsible for encrypting the application traffic before it is transmitted over the channels dedicated to classical communication.
In practice, each encryption module retrieves cryptographic keys from the KM layer and employs these quantum-derived keys to encrypt and decrypt application-layer data exchanged among the APP layer. In doing so, it establishes a proprietary Virtual Private Network (VPN) protocol that is intrinsically interoperable with the QKD layer described above. The interface between the encryption module and the QKD infrastructure is realized through the KM system, and key retrieval procedures conform to the ETSI GS QKD 014 specification \cite{ETSIGSQKD014}.  

The application clients run the Liqo platform, an open-source solution that enables the federation of multiple Kubernetes clusters so that they operate as a single logical cluster.
Within the network, we deploy each application client -- fundamental units of the APP layer -- associated to the cipher of each node, see Fig.~\ref{fig:network}.
The network is designed to ensure quantum-secure data traffic between:
\begin{enumerate}
    \item PoP Milano Caldera and PoP Milano Malpaga;
    \item Malpaga M1 and Malpaga M2 nodes implemented in PoP Milano Malpaga.
\end{enumerate} 
Therefore, the Milano Caldera PoP hosts one ciphers and one application client (Cipher C -- Client C), and
the Milano Malpaga PoP hosts two ciphers and two corresponding application clients (Cipher M1 -- Client M1, Cipher M2 -- Client M2, respectively).

Notice that PoP Milano Data4 is not equipped with ENC and APP units. Indeed, in this network topology, the Data4 node is used as a trusted node, enabling an alternative path connecting PoPs Milano Caldera and Milano Malpaga.

For the sake of clarity, a trusted node is a physically and logically secured site that is allowed to access QKD-generated keys and relay them between adjacent links. In practice, it combines per-link keys and forwards them by one-time-pad (OTP), preserving information-theoretic confidentiality while placing trust in the node.

\section{Methods}
\label{sec:methods}
This section outlines the test plan designed to assess the capabilities of the implemented MAN. In particular, we focus on the QKD and KM layers and on demonstrating that encryption processes remain uninterrupted despite changes in the network configuration, such as strategic optical switching in the QKD layer and simulated Denial-of-Service (DoS) attacks on one of the possible links interconnecting Malpaga and Caldera PoPs.

The primary objective of these tests is to validate the correct integration and functionality of the KM layer within a dynamic QKD-enabled network architecture.
The KM layer is responsible for securely managing and distributing quantum-generated keys to encryption applications, ensuring end-to-end confidentiality across different network configurations and optimal path for key propagation when required. 
Given the complexity introduced by optical switching and trusted-node mechanisms, it is also essential to confirm that the KM system can adapt to changes in the underlying quantum key distribution paths without compromising security or operational continuity.

The motivation for these tests, performed in real life deployment environment, lies in validating the robustness and automation of the QKD network under real operational conditions. 
Therefore, as an initial result, we present the QKD layer performance analysis, with demonstrated stability and flexibility.
By simulating different network topologies and switching events, we ensure that the KM layer maintains secure key distribution regardless of path changes, thereby guaranteeing resilience, interoperability, and compliance with security requirements in a quantum-secure communication environment.
To demonstrate these capabilities, the MAN is tested and evaluated in the following three scenarios:
\begin{enumerate}[label=T\arabic*:, ref=T\arabic*]
\item \label{t1} Direct Caldera -- Malpaga M1 link:
the optical switch activates the direct link between Caldera and Malpaga M1 (Alice1C--Bob1 QKD link) and no keys are present on alternative links. The test demonstrates that the ciphers correctly utilize QKD-derived keys for encryption, generated by the direct QKD link.
\item \label{t2} Caldera -- Malpaga M1 link, via a trusted node in Data4: 
the direct Caldera -- Malpaga M1 link is interrupted simulating a DoS attack. In this scenario, the connection is routed through a trusted node at Data4.
This scenario assesses the system’s ability to maintain secure key provisioning, ensuring that the KMEs can handle multi-hop QKD paths, correctly aggregating keys from separate QKD links (i.e., Alice2--Bob2 and Alice1D--Bob1).
\item \label{t3}
 Inter-Malpaga link and double trusted node:
 with an alternative path switching, the network is dynamically reconfigure to support the inter-Malpaga link (Alice1M--Bob1 QKD link), and the KM layer continues to provide QKD keys to the corresponding ciphers without interruption. In this configuration, both Data4 and Milano Malpaga M1 are used as trusted nodes. 
\end{enumerate}

\section{Results}
\label{sec:results}
The Results section first reports the performance of the QKD links, detailing the secure key rate (SKR), principal performance metric. The SKR is obtained as the ratio between the amount of \emph{secure bits} exchanged and the acquisition time of quantum data needed to generate them.
It is a preferred figure of merit because it quantifies the key-provisioning throughput available to downstream applications.

The results section then presents the outcomes of the three test scenarios introduced in Methods~\ref{sec:methods}. 
We emphasize that, although the QKD and KM layers work cooperatively, they remain conceptually decoupled: the physical layer handles key generation, whereas the management layer is responsible for key storage and distribution.
However, our evaluation and validation of the MAN behavior focus also on verifying the correct, concurrent, and consistent operation of the layers, as well as their interactions across the network.

\subsection{QKD layer performances}
\label{sec:qkd_test}
The implemented MAN demonstrates the capability of the underlying QKD network to generate quantum-secured cryptographic keys in a real-world operational environment, supporting a non-trivial topology which interconnects multiple sites -- using the optical switch or trusted node configurations.
The overall losses experienced by each QKD pair of the network are listed in Table~\ref{tab:loss}. It is worth noting that these values are higher than the channel attenuation measured via OTDR and reported in Table \ref{tab:links} for two main reasons. First, the OTDR trace was taken at 1550 nm, whereas the quantm signals are at 1310 nm, where the attenuation of single-mode fibers is approximately 1.5 times higher. Second, fiber patches used to route light to the racks introduce additional losses, as do the DWDM devices and the optical switch.

The experiments were conducted over an $\sim$14-hour period.
The QKD systems implement the BB84 protocol \cite{bennett2014quantum}, more precisely the 3-state time-bin encoded BB84, with one-decoy method, in the finite-key regime \cite{Ma:2005,tomamichel2012tight, lim2014concise, rusca2018finite}. 

\begin{table}[ht]
    \centering
    \begin{tabular}{lcc}
        \toprule
        \textbf{Link} & \textbf{QKD units} & \textbf{Channel attenuation (dB)} \\
        \midrule
        Malpaga M2 -- Malpaga M1 & Alice1M -- Bob1 & 19.8 \\
        Caldera -- Malpaga M1           & Alice1C -- Bob1 & 21   \\
        Data4 -- Caldera             & Alice2  -- Bob2 &  9.9 \\
        Data4 -- Malpaga M1            & Alice1D -- Bob1 & 26.1 \\
        \bottomrule
    \end{tabular}
     \caption{PoP links, QKD link with involved devices and overall channel attenuation.}
     \label{tab:loss}
\end{table}

\begin{figure}[t]
\centering
    \begin{subfigure}[b]{0.45\textwidth}
        \includegraphics[width=1\linewidth]{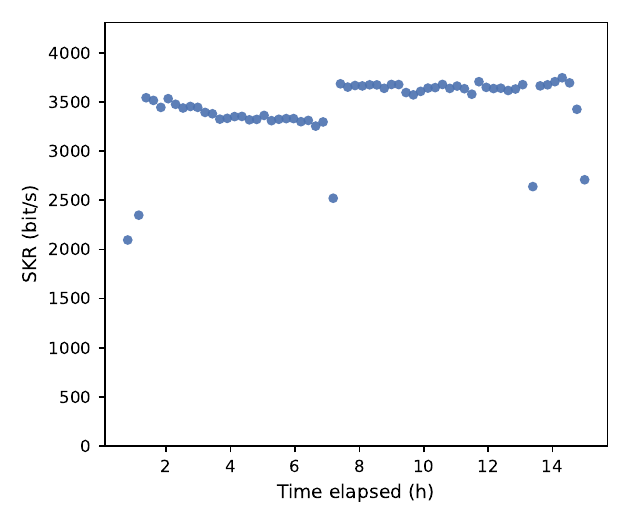}
        \caption{Point-to-point link}
        \label{fig:skr_sta}
    \end{subfigure}
    \begin{subfigure}[b]{0.45\textwidth}
        \includegraphics[width=1\linewidth]{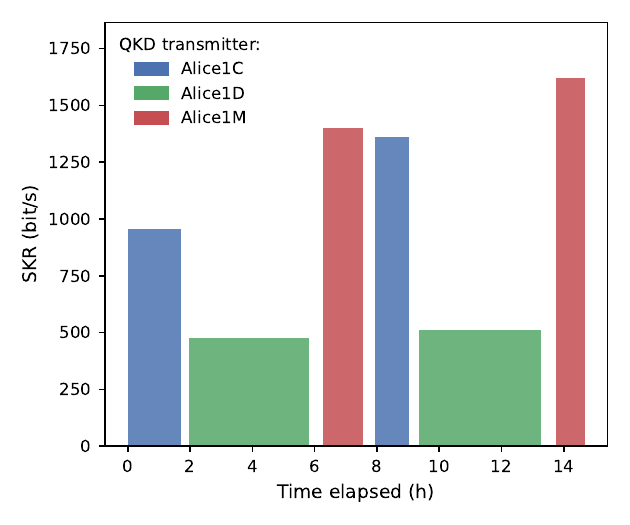}
        \caption{Star configuration, enabled by the optical switching  }
        \label{fig:skr_switch}
    \end{subfigure}
    \caption{Secure key rate (SKR) of the QKD network during the field trial. \subref{fig:skr_sta}) shows the performance of the point-to-point link between the Alice2 and Bob2 QKD devices, whereas \subref{fig:skr_switch}) refers to the switched links (Alice1M in node Malpaga M2, Alice1C in node Caldera, Alice1D in node Data4, and Bob1 in node Malpaga M1). In the latter plot, the height of each bar represents the mean SKR during the active period of the corresponding link, while its width reflects the time duration of the specific active configuration.}
\end{figure}

\paragraph{Point-to-point QKD link.}
The link between Caldera and Data4 PoPs is the only connection not routed through the optical switch.
The SKR provided by the QKD link, shown in Fig.~\ref{fig:skr_sta}, remains stable over  
the whole testing period and is consistent with the expected values for a 10\,dB link. 
Periodically, the system performs a recalibration phase to optimize performance and to adapt to drifts in the environmental conditions. In the SKR trace, shown in Fig.~\ref{fig:skr_sta}, this behavior can be noticed by the low-SKR points, which are caused by the additional time required by the calibration phase. 

\paragraph{Optical switch.} 
The remaining links of the QKD layer are arranged in a star configuration, with the receiver Bob1 as the center, see Fig.~\ref{fig:network}. Through the optical switch, located in PoP Milano Malpaga and controlled by the KM layer, Bob1 can exchange keys with the three different transmitter units: Alice1M, Alice1C, and Alice1D.

Throughout the demonstration, the optical switch sequentially routed the connection among the transmitters, and thus the corresponding paths. This is reflected in the SKR histogram in Fig.~\ref{fig:skr_switch}, where active-link intervals are color-coded blue for Alice1C--Bob1, red for Alice1M--Bob1, and green for Alice1D--Bob1, each indicating the period during which that link was selected by the switch.\\
The KM layer controls the switching mechanism by monitoring the flow of generated keys and dynamically changing the communication path at strategic intervals.
This switching process is driven by the system’s objective of maintaining a balanced key distribution across all paths. By continuously monitoring the key inventory at each node, the SDN controller determines when a path switch is required, ensuring that the difference in key counts remains below a predefined threshold. 
This behavior is evident from Fig.~\ref{fig:skr_switch} and is further reinforced by the observation that a more efficient link remains active for a shorter duration, as it reaches the predefined threshold significantly faster than a link operating at a lower key generation rate. During this test, the threshold is configured to ensure that the difference in the number of available keys across all links remains below 15000 cryptographic keys, each with a length of 256 bits.

Overall, the results demonstrate stable and reliable performance across multiple switching events.

\subsection{Results on the Key Management and Encryption Layers }
\label{sec:KM_ENC_test}
The KM layer plays a critical role in ensuring secure and efficient distribution of quantum-generated keys to encryption applications within a QKD-enabled network. 
Given the dynamic nature of the network, where an optical switch and trusted node configuration introduce variability in key distribution paths, it is essential to validate the robustness and adaptability of the KM layer under real operational conditions.

During the optical-link switching phase, two Internet Control Message Protocol (ICMP) exchanges are initiated: one between the nodes in Malpaga and another one between Malpaga node and the Caldera node. ICMP is employed because continuous ping sessions provide a simple and reliable method for monitoring connectivity and detecting even brief packet-loss events.
In particular, the results show that throughout the optical switching operation on the trusted node, no packet loss or communication interruption occurred. This confirms that the connection between each cipher and its associated KME remains stable and fully operational during the re-keying process.\\

Three distinct scenarios have been defined in Section~\ref{sec:methods} to evaluate the KM layer and the encryption layer response. 
Refer to Fig. \ref{fig:network} to correspondence among KMEs and ciphers. Note that there is no cipher present in Data4 since, in this demonstration, this location always acts as a trusted node. Therefore, KME D does not have an associated cipher. 

\subsubsection{Test 1: Direct Caldera -- Malpaga M1 QKD link}
\label{sec:test1}
In the first Test \ref{t1}, the configuration entailed activation of the optical switch on the Alice1C transmitter, thus targeting a direct quantum channel between the Caldera (Alice1C) and Malpaga M1 (Bob1) sites while excluding any alternative transmitters (Alice1D, Alice1M) and respective paths.
\paragraph{Validation of the QKD layer}
The evolution of the SKR is shown over time, in Fig.~\ref{fig:skr_switch}. The active QKD link is colour-coded by network configuration, each one being determined by the optical-switch routing, as detailed in Section~\ref{sec:methods}.
The first active path, the direct Alice1C--Bob1 link (blue bar) is selected on the optical switch, and the system generates keys between Caldera and Malpaga. The corresponding SKR samples appear in the initial blue-coded bar of the histogram in Fig.~\ref{fig:skr_switch}.

System logs consistently report the activation of the routing of the optical switch to Alice1C transmitter. Coherently, the QKD system reports the deactivation of the remaining switch ports, i.e. the optical paths to transmitters Alice1M and Alice1D.
Taken together, the temporal alignment between the switch state and the appearance of SKR provides a direct validation that quantum signals traverse the intended optical path and that the QKD device is operating nominally.
Key generation proceeds as soon as the Alice1C--Bob1 path is activated and persists until the ensuing reconfiguration event.

\paragraph{Validation of the KM layer.} 
For the KM plane, the test objective is to demonstrate that the two involved KM units (KME M1 and KME C) receive requests from their corresponding encryption devices (Cipher M1 and Cipher C, respectively), without involving the trusted node located at Data4 (KME D).
From the collected logs, KME M1 (Malpaga M1) shows that the local encryption device, Cipher M1, initiates a key request addressed to its remote counterpart, Cipher C. In parallel, KME C (Caldera) documents the complementary request originating from its own local encryption device, Cipher C,  and directed toward the remote Cipher M1.\\
Crucially, KME D (Data4 trusted node) shows no evidence of key propagation toward any remote KME responder. This unambiguously excludes the trusted node from the exchange and is consistent with a direct, point-to-point Caldera -- Malpaga M1  session over the Alice1C-Bob1 quantum link.\\

We therefore operationally validate that the ciphers at Caldera and Malpaga are supplied with QKD-derived keys generated on the direct QKD link Alice1C--Bob1, without transiting through the Data4 trusted relay.

\subsubsection{Test 2: Caldera -- Malpaga M1 linkk,  via Data4 as trusted node}
The network features two distinct paths connecting Caldera and Malpaga.
The direct link, Alice1C--Bob1, has been detailed and validated in Test \ref{t1}.
Test \ref{t2} targets the alternative, two-hop path that traverses the Data4 PoP acting as a trusted node.
In this configuration, Data4 concatenates the Alice2--Bob2 QKD segment (Caldera -- Data4) with the Alice1D--Bob1 QKD segment (Data4 -- Malpaga M1),  enabling the key-establishment between the two PoPs.  
Operationally, the optical switch activates the Alice1D--Bob1 QKD link, and disables all remaining routes. 
The KM layer then serves the ciphers by aggregating the keys generated by the two segments.

\paragraph{Validation of the QKD layer}
In the SKR histogram of Fig.~\ref{fig:skr_switch}, the reconfiguration that selects the Alice1D transmitter is visible and  highlighted by the color change of the second bar. 
Indeed, the second active path, whose key generation is scored in green, corresponds to the activation of the Alice1D--Bob1 QKD link, persisting until the next network configuration. 
The QKD system logs corroborate the activation of this link, connecting Malpaga M1 to Data4, and the concurrent deactivation of the Alice1M and Alice1 C ports.

\paragraph{Validation of the KM layer}
The activation of Alice1D--Bob1 (Data4 -- Malpaga M1), combined with the running Alice2--Bob2 system (Caldera -- Data4) enables the alternative Caldera -- Malpaga M1 path via the Data4 trusted node. 
To preclude inadvertent use of residual direct-link keys, we explicitly flush the key buffers shared by KME M1 (Malpaga M1) and KME C (Caldera), thereby emulating a DoS-like exhaustion on the direct path.
As a result, the KM system is compelled to route cipher requests through the trusted node, KME D (Data4).\\
Under this condition, KME M1 (Malpaga M1) records that local encryption device, Cipher M1, initiates a key request targeting the remote peer Cipher C; symmetrically, KME C (Caldera) records the complementary request generated by its local encryption device, Cipher C, for the remote Cipher M1. Crucially, and in contrast to Test \ref{t1}, KME D (Data4) logs evidence inter-KME key propagation across the trusted node. This unambiguously indicates that key provisioning between the endpoints transits through the trusted node rather than being served exclusively by local key pools. \\

The concordant evidence of a non-zero SKR coincident with activation of the Alice1D--Bob1 link and the expected KM layer behavior operationally validates that Cipher C at Caldera and Cipher M1 in Malpaga M1 operate using QKD-keys provisioned via the two-hop QKD path passing from Data4, confirming the trusted node functionality.

\subsubsection{Test 3: : Inter-Malpaga and double trusted node}
In the third and final test \ref{t3}, the optical switch is set to activate the Alice1M--Bob1 QKD link and disable the remaining paths. 
The final test results in a multi-hop network configuration with two operating trusted nodes. 
Indeed, the network is configured now to enable key establishment from Caldera (Cipher C) to the simulated external node implemented in Malpaga M2 (Cipher M2), exploiting both trusted nodes: Data4 and Malpaga M1. 

This scenario therefore extends Test \ref{t2} by introducing a second trusted node. Operationally, the active QKD segments and conditions are the following: 
\begin{enumerate}
    \item {Alice2--Bob2:} continuously running Caldera -- Data4 point-to-point system;
    \item	{Alice1D--Bob1:} Data4 -- Malpaga M1 link with keys already established in support of the Caldera -- Malpaga M1 connection of Test \ref{t2};
    \item	{Alice1M--Bob1:} inter-Malpaga link, activated by the optical switch.
\end{enumerate}

\paragraph{Validation of the QKD layer}
In Fig.~\ref{fig:skr_switch}, the third active path appears in red and corresponds to the Alice1M--Bob1 selection. 
The QKD system logs confirm the proper setting of the optical switch, routing connection inside the Malpaga PoP, and that the two alternative routes, Bob1--Alice1C and Bob1--Alice1D, are disabled.

\paragraph{Validation of the KME layer}
The activation of the Alice1M--Bob1 QKD link, with consequent key generation, combined with the key material produced during the Test \ref{t2} configuration (Caldera -- Malpaga M1, with trusted node in Data4) finally allows the connection between Caldera and the Malpaga M2 node, through two trusted nodes.
Consistent with a correct association between these endpoints, the systems give evidence that at KME M2 (Malpaga M2), a key request is issued by the local encryption device, Cipher M2,  targeting the remote Cipher C,
and that at KME C (Caldera), the complementary request is generated by its local encryption device, Cipher C, for the remote Cipher M2.\\
Critically, in line with the extension over Test \ref{t2}, both 
KME D (Data4) and KME M1 (Malpaga M1) simultaneously exhibit invocations of key propagation.
This constitutes positive evidence that key provisioning transits through \emph{both} trusted nodes (Data4 and Malpaga M1), rather than being served locally at either endpoint.\\
The described indicators validate correct operation of a QKD network employing two trusted nodes. In this configuration, Bob1 generates secure keys with Alice1M, supporting the internal Malpaga link. Combining the resulting key material with keys established over Caldera -- Data4 -- Malpaga M1 chain from Test \ref{t2}, Cipher C (Caldera) can finally operate with Cipher M2 (Malpaga M2).\\

The three test cases, \ref{t1}, \ref{t2} and \ref{t3}, reported above correspond to the first three color-coded reconfiguration bars in the SKR histogram of Fig.~\ref{fig:skr_switch}, i.e., the initial activation windows associated with the Alice1C--Bob1, Alice1D--Bob1, and Alice1M--Bob1 links, respectively. 
After these initial intervals, the QKD-KM architecture continues to switch among the Malpaga star node as described in Section~\ref{sec:qkd_test} dynamically selecting the active path to maintain a balanced key inventory across links. 

\subsection{Results on the applicative layer}
\label{sec:test_appl}
The application layer, developed by Politecnico di Torino, provides the end-user services that generate the data traffic protected across the quantum-secure network. This layer is built upon the Liqo open-source platform, which enables dynamic and seamless federation of multiple Kubernetes clusters. In the field trial, this architecture allowed heterogeneous nodes located at the Milano Malpaga (M1, M2) and Milano Caldera (C) PoPs to operate as a single logical cluster, with Liqo acting as a load balancer for distributed services.

To validate the performance and reliability of the integrated communication stack, a comprehensive test suite was executed using three primary methodologies: high-intensity TCP throughput measurement via Iperf, large-scale file transfers (approx. 2.5 GB) using SFTP, and HTTP load testing with Apache Benchmark (ab) on a cluster-wide web server. These evaluations were conducted across different node pairs, (M1, C), (M1, M2), and (C, M2), to assess the system's behavior under realistic operational workloads .

Overall, the application-layer validation confirms that the quantum-secure network can support high-intensity data traffic without significant performance degradation. These findings demonstrate the functional maturity and operational readiness of the entire stack, proving it is capable of handling real-world metropolitan-scale communication scenarios.

\section{Conclusions}
\label{sec:conclusions}
We reported the field deployment and end-to-end validation of a metropolitan-scale quantum-secure network interconnecting multiple data-center locations in the Milan area. The infrastructure integrates QKD hardware, a standards-compliant Key Management layer, and a centralized SDN-based control plane, operating over installed dark fibers with active optical switching and trusted-node routing. 
This layered design enabled automated path orchestration, inventory-aware key balancing, and seamless key consumption by applications, demonstrating operational readiness under realistic conditions.

The test suite validates resilience and functional correctness across three representative scenarios. With the direct Caldera -- Malpaga M1 quantum path (T1), ciphers consume QKD-derived keys as expected. 
Under a simulated Denial-of-Service on that link, the system preserves uninterrupted key provisioning by rerouting via the trusted node at Data4 (T2), correctly aggregating per-link keys along the multi-hop path. 
Finally, when reconfigure to sustain the inter-Malpaga link with a double trusted-node arrangement (T3), the KM layer continues to serve keys without disruption. 
Across all cases, orchestration and automation maintain service continuity, evidencing the maturity of the overall integration.

From a networking perspective, the trial highlights the feasibility of operating a complex metropolitan QKD network comprising a star sub-topology and trusted-node routing on top of a existing fiber plant, with centralized SDN control and optical switching. 
The ability to reconfigure quantum paths on demand, while enforcing inventory targets and keeping applications agnostic to the underlying changes, is a critical step toward scalable, cost-effective deployments in large operators’ networks.

Looking ahead, the architecture is directly extensible to additional nodes. 
Future work will target long-term field operation under continuous mixed traffic, consolidating the reference architecture presented here for real-world quantum-secure metropolitan networks.

\section{Data availability statement}
The data generated during and/or analysed during the current study are available from the corresponding author on reasonable request.

\section{Conflict of interest statement}
The authors declare no conflicts of interest.

\section{Acknowledgments}
The Authors acknowledge financial support from EQUO (European QUantum ecOsystems) project, funded by the European Commission in the Digital Europe Programme [GA No. 101091561]. DB acknowledges support from the European Union ERC StG, QOMUNE, 101077917. AZ acknowledges support from the European Union - NextGeneration EU, "Integrated infrastructure initiative in Photonic and Quantum Sciences" - I-PHOQS [IR0000016, ID D2B8D520, CUP B53C22001750006]. 

\bibliography{biblioMAN}

\end{document}